# Two Novel Server-Side Attacks against Log File in Shared Web Hosting Servers

Seyed Ali Mirheidari[1], Sajjad Arshad[2], Saeidreza Khoshkdahan[3], Rasool Jalili[4]

[1]Computer Engineering Department, Sharif University of Technology, International Campus, Iran
[2]School of Electrical and Computer Engineering, Shahid Beheshti University, General Campus, Iran
[3]Sabzfaam Information Technology Corporation, Iran
[4]Computer Engineering Department, Sharif University of Technology, Iran

mirheidari@kish.sharif.edu, s.arshad@mail.sbu.ac.ir, khoshkdahan@sabzfaam.ir, jalili@sharif.edu

*Abstract*—**Shared Web Hosting service enables hosting multitude of websites on a single powerful server. It is a well-known solution as many people share the overall cost of server maintenance and also, website owners do not need to deal with administration issues is not necessary for website owners. In this paper, we illustrate how shared web hosting service works and demonstrate the security weaknesses rise due to the lack of proper isolation between different websites, hosted on the same server. We exhibit two new server-side attacks against the log file whose objectives are revealing information of other hosted websites which are considered to be private and arranging other complex attacks. In the absence of isolated log files among websites, an attacker controlling a website can inspect and manipulate contents of the log file. These attacks enable an attacker to disclose file and directory structure of other websites and launch other sorts of attacks. Finally, we propose several countermeasures to secure shared web hosting servers against the two attacks subsequent to the separation of log files for each website.**

## I. INTRODUCTION

Nowadays with Internet popularity increasing, many people are creating their own websites. In order to publish, many people prepare their own dedicated servers. But with the increase in hardware power, it is possible to host several websites on a single physical server. This kind of web hosting is commonly known as shared web hosting. In shared web hosting, the physical resources are shared among different websites simultaneously. Also, the administration of the webserver is handled by web hosting providers and owners of these websites do not need much information and experiences about the administration of their websites. However, limited resources force users to suffer from low performance. Furthermore, shared web hosting servers have some security issues since there is no proper isolation between different websites [1]. According to the Zone-H site, the world has witnessed an increasing number of website defacements [2]. In other words, websites that are co-located with a vulnerable website on a physical server might be in danger too and a noticeable number of defacements are released in only one IP or physical server mass deface [3,4].

In this paper, we introduce two novel server-side attacks against the log management system of webservers in shared web hosting servers: Log Poisoning and Log Snooping. In order to be vulnerable to these attacks, a webserver must be setup with default configuration in respect to how logs are stored, thus an attacker who controls a website hosted on a shared web hosting server is able to attack all other websites hosted on the same webserver. In other words, he is able to manipulate logs of other websites (Log Poisoning) or to inspect their logs (Log Snooping). This way, an attacker can steal private information, reveal file and directory structure of websites, and use these attacks to launch other complex attacks.

In this paper, we focus on the Apache webserver to present the attacks. According to Netcraft [5], Apache webserver has the highest usage among other webservers such as Microsoft IIS. Since most countermeasures are developed for POSIX operating systems, this study mainly focuses on Linux operating system. Also, we use PHP programming language because of higher popularity, usability and reliability. However, the discussed attacks are not unique to Apache webserver and every webserver installed with certain configuration is potentially vulnerable to aforementioned attacks.

The rest of this paper is structured as follows: In Section II the overall architecture of shared web hosting servers is portrayed. We describe two novel server-side attacks against log file in Section III. In Section IV, we present other complex attacks which can be launched as a result of two aforementioned server-side attacks. In Section V, we present several countermeasures against server-side attacks and we conclude our paper in Section VI.

## II. SHARED WEB HOSTING ARCHITECTURE

In this section we discuss the details of shared web hosting architecture to obtain a better view for understanding the attacks presented in Section III. In shared web hosting, a webserver is hosting many websites, simultaneously. The website owner has a FTP account which can upload new files for his website and uploaded files are owned by the owner user account. A webserver run as a specific user account (apache, daemon, and www-data) and handles all HTTP requests for all websites. So, webserver must be able to read the files on each website. However, in some Content Management Systems (CMS), the users must be able to upload files and therefore webserver needs the write access to website directories besides read access. Fig. 1 depicts the necessary permissions for Apache webserver in Linux operating system where web1 and web2 are Linux users and owners of two different websites hosted on the shared web hosting server.

```
d rwx r-x ---    web1:www-data    /var/www/site1
d rwx r-x ---    web2:www-data    /var/www/site2
```

Figure 1. Essential Permissions for Apache Webserver

In shared web hosting, there are two general forms of webserver configuration for executing) scripts as below:
- Configure webserver to load the script interpreter as a webserver module.
- Configure webserver to run the script interpreter as a CGI binary.

A webserver module is loaded by webserver process or is compiled into the webserver binary, which means the webserver process contains a binary image of the interpreter. A CGI is executed as a single process for each request, meaning that the webserver will create a new interpreting process for each arriving request. Using script interpreter as webserver module is more stable under load and much more efficient in handling requests and resource management, than the CGI mode. But CGI mode is more secure because malicious scripts do not affect webserver process.

### III. LOG FILE ATTACKS

Web servers usually store the information of processed requests in a log file. A log file usually includes some information like Domain Name, Client IP, Request Time, Request Type (GET or POST), Requested Filename, and Size of Transferred File and Return Status Code from webserver. The two attacks presenting are based on this fact that webserver uses a single file for storing logs of various websites and the log file is accessible by every script executed by the webserver. These weaknesses enable an attacker to open log file in a write mode and modify logs residing on the same webserver (Log Poisoning). It also allows the attacker scripts to open log file, inspect logs of other websites and misuse the information that is supposed to be private (Log Snooping). The details of these attacks are presented in the following two sections.

#### A. Log Poisoning

In shared web hosting with default configuration, log file can be modified only by the root user and is only readable by other users. On the other hand, webserver should have a permission to write in log file, regardless of the user account that is run with. Therefore, in most webservers like Apache, parent webserver is executed with root privilege and child webservers are run by parent webserver to handle the requests. In some operating systems like Linux, file descriptors opened by parent process, will be inherited to child processes. This way, parent webserver can open the log file in write mode and fork child webservers to allow them to write in log file. Thus, log file descriptor is inherited by child webservers and consequently they can modify log file although they are not run with root privilege. Since scripts of websites hosted on the shared web hosting server are usually executed by child webserver processes, they are able to modify the log file. In Log Poisoning attack, an attacker creates a script to find log file descriptor and open the log file in write mode. For instance, in Linux operating system, information about open files of each process exists in */proc/PID/fd*, in which PID is the process ID. Then, an attacker creates a PHP script to find the open files of child webserver process which executes the script and re-opens the log file with write access. The sample PHP script for Log Poisoning attack is shown in Fig. 2.

To be susceptible to this attack, Apache must use PHP interpreter as an Apache module because when Apache runs PHP interpreter as CGI, the new PHP interpreter process does not inherit log file descriptor from Apache, so the malicious PHP script is not able to re-open log file with write access and modify its content.

Having the write access to the log file, attackers can do malicious tasks like clearing other website requests in order to cover track of their penetration or adding some fake requests to the log file. Generally, it is true to say that write access to log file in shared web hosting servers has harmful consequences and attackers can accomplish various attacks on victim websites by poisoning the log file.

#### B. Log Snooping

In default configuration, log file is readable by all users. So, webserver user can read the log file and consequently all scripts run by the webserver are able to read the contents of log file. Therefore, scripts of a website are able to read the logs of other websites located on the same shared web hosting server. In Log Snooping attack, an attacker searches the victim website logs to retrieve important information and use the information to follow malicious activities. Unlike Log Poisoning attack, Log Snooping attack is feasible in two modes which webserver runs the script interpreter (Module or CGI).

```php
<?php
    if ($dh = opendir('/proc/self/fd/')) {
        while (($fd = readdir($dh)) !== false) {
            if (strpos(realpath($dir.$fd), "access_log") !== false) {
                $log_fd = $fd;
                break;
            }
        }
        closedir($dh);
    }
    $file = fopen("php://fd/$log_fd", "w");
    fwrite($file, "Some Junk Data\n");
    ...
    fclose($file);
?>
```

Figure 2. Log Poisoning Script (PHP-Module Mode)

```php
<?php
    if ($dh = opendir('/proc/self/fd/')) {
        while (($fd = readdir($dh)) !== false) {
            if (strpos(realpath($dir.$fd), "access_log") !== false) {
                $log_fd = $fd;
                break;
            }
        }
        closedir($dh);
    }
    $file = fopen("php://fd/$log_fd", "r");
    $data = fgets($file);
    ...
    fclose($file);
?>
```

Figure 3. Log Snooping Script (PHP-Module Mode)

```php
<?php
    $file = fopen("/var/log/apache/access_log", "r");
    $data = fgets($file);
    ...
    fclose($file);
?>
```

Figure 4. Log Snooping Script (PHP-CGI Mode)

If the webserver administrator made the log file unreadable for other users, attacker can use the PHP script shown in Fig. 3 to accomplish Log Snooping attack. Also, Fig. 4 access to the log file enables attackers to access much useful information. One of the most important information is the structure of files and folders of victim websites. Attacker can reconstruct the site tree using the requested URLs and be informed about the names of website files and folders. For example, in several hardening best practices, the name of administrator authentication page is changed in order to prevent the attackers from entering administration panel. But with using site tree, attacker can bypass this technique and find the authentication page. Then the attacker can use techniques like SQL Injection to obtain hashed password of administrator and find clear password text by using brute force of encoded password or using brute force for both user and password in order to obtain admin login credentials. It is important to know that if the attacker does not have access to the shared web hosting server, he will not be able to find the authentication page easily.

## IV. RESULTING ATTACKS

The Log Poisoning and Log Snooping attacks can be used as intermediate steps to accomplish other attacks against the websites residing on a shared web hosting server. The resulting attacks will be presented in the following sections with more details.

### A. Executing Malicious Code

Log Poisoning attack enables attackers to execute malicious codes with vulnerable website rights. Some websites are vulnerable since they allow special code reuse by including files. In other words, users supply the values of some parameters used in URL in order to include desired files. In this case, attackers try to misuse and include some malicious files. One of the most common attacks in this area is known as Local File Inclusion (LFI) [6] which leads in including victim server local files [7]. During recent years, several methods such as LFI2RCE [8,9] are proposed which are able to execute remote code using LFI attack. One of such methods is adding some malicious code to the log file of webserver and including the log file by LFI which leads in execution of malicious code by victim website. However, without having access to the local victim file system, poisoning the log file is a complicated and hard task and, sometimes impossible. In shared web hosting servers, using Log Poisoning attack, an attacker can add some malicious code to the log file easily and accomplish the LFI2RCE attack.

### B. Drawing Site Tree

As mentioned before, if an attacker reads the webserver log file, he will be able to draw the site tree which includes file and directory names of different websites and use this information in dangerous ways. In many web vulnerability scanners, crawling is definitely the most important part due to this that the scanner might miss vulnerabilities. So if the crawling engine is weak, the scanner will certainly miss the vulnerabilities [10]. If an intruder has access to victim webserver log file, he can draw an accurate site tree like as web vulnerability and pass this arguably phase. In other words, accurate site tree is the first step of successful penetration testing cycle.

### C. Finding Co-located Websites

Depending on the configuration of webserver, there are various methods for identifying the websites hosted on the shared web hosting server. The attacker can write a script to list directories recursively and finds the names of co-located websites. Also, using Log Snooping attack, attacker is able to find co-located websites. As mentioned before, there is a variety of information about client requests available in the log file. One is the virtual host name or website domain name that is serving the request. Hence, the attacker can read the content of log file and list the names of websites that are located on the server.

### D. Revealing Sensitive Information

Generally most developers send authentication tokens (usernames, passwords, session identifiers) via GET variables and because webserver records information provided by GET variables, sensitive information reveals in case of Log Snooping. For instance, consider the below URL which is sent to the victim website, once the submit button on authentication page is clicked:

*https://www.victim.com/login?user=admin&pass=plain_or_hash_pass*

Most webservers with default configuration log this GET request with related parameters in clear text. Therefore, an intruder can use this information to login to the victim website as a valid user.

## V. COUNTERMEASURES

Since using shared web hosting is popular, securing the solution is a more proper idea than skipping this service. For this purpose, several methods have been proposed in order to make a more secure shared web hosting installation [11,12,13,14,15]. In this section, we present countermeasures developed for Apache webserver on Linux and how it confronts the server-side attacks described in Section III. If we examine these attacks carefully, we can figure out the main cause of the attacks is the lack of proper isolation between log files of different websites.

In default configuration, webservers use a single log file for recording request of all websites hosted on the shared web hosting server. In order to stop exampled attacks against log file, it is a common practice to create separate log file for each website and put log files in separate directories [16]. A sample configuration in Apache webserver for creating separate log file for each virtual host website is shown in Fig. 5. In addition, the proper permissions must be set on the log file directories as the malicious scripts cannot read from or write on them. The necessary permissions on log file directories in Linux operating system are depicted in Fig. 6. In Fig. 6, web1 and web2 are owners of the corresponding websites.

In default configuration, Apache is executed by a unique user who has access to every website. A guessable idea is that Apache runs each website by its owner user account. Therefore, different methods have been introduced for Apache in the past years. As the first attempts, suEXEC [17] and suPHP [18] have been introduced as Apache modules. The suEXEC is a wrapper binary file and an Apache module. When a HTTP request arrives, Apache runs the wrapper and finds the script name and User/Group ID [1]. This module can only be used with CGI [19] or FastCGI [20] programs.

```
<VirtualHost *:80>
    DocumentRoot      /home/website1/public_html
    ServerName        website1
    ErrorLog          /home/website1/log/error_log
    CustomLog         /home/website1/log/access_log common
    ...
</VirtualHost>

<VirtualHost *:80>
    DocumentRoot      /home/website2
    ServerName        website2
    ErrorLog          /home/website2/log/error_log
    CustomLog         /home/website2/log/access_log common
    ...
</VirtualHost>
```

Figure 5. Log Separation in Apache for Each Website

```
d rwx r-x ---    web1:web1    /home/website1
d rwx r-x ---    web1:web1    /home/website1/public_html
d rwx r-x ---    web1:web1    /home/website1/log

d rwx r-x ---    web2:web2    /home/website2
d rwx r-x ---    web2:web2    /home/website2/public_html
d rwx r-x ---    web2:web2    /home/website2/log
```

Figure 6. Necessary Permissions for Log Files' Directories in Linux

In order to install suEXEC, you must prepare a unique CGI or FastCGI binary file for each website and user and group ID of the owner must be set as owner of website. To be mentioned, using suEXEC with CGI has very low performance in a way that Chary has named it as a performance killer [12]. Same as suEXEC, suPHP runs PHP scripts with the specified user and group ID. In contrast to suEXEC, there is no need of a unique CGI or FastCGI binary file for each website with suPHP module. Also as same as suEXEC, suPHP suffers from low performance [1].

When Apache 2.0 has been released, different MPM [21] methods have been introduced. Some of them are developed to solve the shared web hosting security problem. Sean Gabriel Heacock introduced Peruser MPM [22]. Peruser MPM uses processes instead of threads to handle requests. This MPM runs a control Apache process as root privilege and the control process creates several multiplexer processes with Apache user privilege. The multiplexer process listens on port 80, accepts incoming requests and reads the request to check, from destination website. Then, it passes the request to relevant worker process to handle it. The worker processes run under the user and group ID of respective website owners. Also the control process always maintains a pool of idle worker processes to increase the performance [1].

Another relevant MPM introduced by Steinar Gunderson is ITK MPM [23]. ITK MPM creates a managing Apache process with Root privilege. The managing process spawns several listeners with root privilege. The listener process listen on port 80 and handles new request to determine which website it is. Then, it creates a new Apache handler process with user and group ID of website owner to serve the request. But, the main difference of ITK MPM with Peruser MPM is that after the request has been completed, the handler Apache process is terminated. In other words, ITK MPM does not maintain a pool of idle handler processes for serving the requests. Due to this, ITK MPM is a good solution, if the server has high number of users.

According to [15], the ITK MPM solution behaves relatively well in all aspects. However, allocating separate log file for each website does not seem a perfect solution all the time, since by increasing number of websites; it will cause some problems with insufficient file descriptors.

VI. CONCLUSION

Shared web hosting is the most common type of web hosting due to its low monthly costs and the need of almost no knowledge and experience from the customer side for administration of their websites. However, the websites hosted on the shared web hosting servers suffer from some security weaknesses.

This paper addressed two novel server-side attacks which exploit the lack of proper isolation between the log files of different websites resided on a shared web hosting server. We demonstrated that webservers using a single log file to store website logs are prone to an attacker in control of a website hosted on a shared web hosting server can manipulate and inspect logs of other websites hosted on the same server, thus the attacker is able to steal private information, reveal file and directory structures of other websites and conduct other complex attacks. Eventually, we presented countermeasures and how they secure the shared web hosting installations.